# Atmospheric electricity and thunderstorm ground enhancements

A.Chilingarian[1,2,3], G. Hovsepyan[1,2], M.Zazyan[1,2]

[1]A. Alikhanyan National Lab (Yerevan Physics Institute), Yerevan 0036, Armenia

[2] HSE University, Moscow 109028, Russia

[3]National Research Nuclear University MEPhI, Moscow 115409, Russia

## Abstract

The comparative analysis of three thunderstorms on Aragats in May 2021 demonstrates that relativistic runaway electron avalanches (RREAs) are developing in large areas of the thunderous atmosphere. In the active storm zone, RREAs last tens of seconds to a few minutes, until lightning flashes terminate electron acceleration. If the lightning activity is far from the detector site, the measured enhancements of particle fluxes (thunderstorm ground enhancements - TGEs) smoothly decay when the atmospheric conditions cannot anymore sustain the electron runaway process. In this case, the TGE has a more or less symmetrical shape and can last up to 10 minutes and more. Thus, RREAs development is paired with lightning activity, creating huge electron fluxes preceding the development of lightning leaders. We show that the total surface area exposed to ionizing radiation can reach 100 km$^2$ and the total number of gamma rays directed to the earth's surface – can be estimated as $\approx 1.3*10^{16}$ (for TGE particles with energies above 300 KeV). The differential energy spectra of electrons and gamma rays recovered from the data of particle spectrometers are used to estimate the height of a strong accelerating electric field region, which can extend down to tens meters above the earth's surface.

## Plain language summary

Many species of elementary particles are born in the terrestrial atmosphere by high-energy protons and fully-stripped nuclei accelerated at exotic galactic sources. During thunderstorms, in addition to this more-or-less constant flux, electrons and gamma rays are produced by the most powerful natural electron accelerator operated in the electrifying atmosphere. Huge fluxes of electrons and gamma rays can exceed the background up to 100 times and pose yet not estimated influence on the climate. More than 2,000 thunderstorms are active throughout the world at a given moment, producing on the order of 100 flashes per second. The overall surface of the thunderous atmosphere each moment can be estimated as $\approx$ 200,000 km$^2$, and according to our estimates $\approx 1.3*10^{16}$ gamma rays with energies above 300 keV are hitting the earth's surface each second. The long-term effects of this radiation on humans should be thoroughly examined.

## Key points

- The most powerful natural electron accelerators operated in thunderclouds around the globe send $\approx 1.3*10^{16}$ high-energy particles with energies above 300 keV in direction of the earth's surface each second;
- The TGE electrons are precursors of the lightning flashes by opening ionized channels for the lightning leaders;
- The strong accelerating electric field can extend down to a few tens meters above the earth's surface

1. Introduction

It is widely accepted, that the cloud charge structure for a typical thunderstorm contains an upper positive charge region consisting of ice crystals, a main negative charge region consisting of both graupel and ice crystals, and a lower positive charge region consisting of graupel (Kuettener, 1950). The electric charge of graupel is positive at temperatures warmer than -10° C, and negative at temperatures cooler than -10° C (Takahashi, 1978, Wada et al., 2021). In review (Williams, 1989) was stated that the tripolar structure of thunderstorms is supported by a wide variety of observations and that temperature appears to be the most important single parameter in controlling the polarity of charge acquired by the precipitation particles. When graupel falls into the region warmer than ≈ -10° C, a charge reversal will occur in the central part of the storm, and the graupel population will change the charge from negative to positive. Large and dense graupel population either suspended in the middle of the thunderstorm cloud or falling toward the earth's surface constitutes a "moving" lower positive charge region (LPCR). The dipole formed by the LPCR and relatively stable main negative (MN) charge region significantly intensify the electric field of the dipole formed by the MN and its mirror image in the ground (MN-MIRR, first scenario of RREA initiation, see Fig.1 in Chilingarian et al., 2021a). A free electron entering the strong and extended electric field accelerates and unleashes the relativistic runaway electron avalanches (RREA, Gurevich et al., 1992). The RREA is a threshold process, which occurred only if the electric field exceeds the critical value in a region of the vertical extent of about 1–2 km. When the second scenario of the RREA origination (MN-MIRR plus MN-LPCR, see the details of the RREA initiation model in Chilingarian et al, 2020 and 2021a) is realized the electric field in the cloud frequently surpasses the critical value and an intense RREA ends up in an extreme thunderstorm ground enhancement (TGE, Chilingarian et al., 2010, 2011) sometimes exceeding the background level of gamma rays and electrons up to hundred times (Chum et al., 2021). After the graupel fall, or after the lightning flash neutralizing the positive charge of the LPCR, the surface electric field again is controlled by the main negative charge region only.

At Aragats research station we develop a new approach for understanding RREA development in the strong electric fields, and after exiting these fields. 24/7 monitoring of the TGE particle fluxes and energy spectra allows to understand the charge structure of the lower part of the thundercloud and its dynamic changes. The most pronounced consequence of the complex interplay of TGEs and lightning activity observed now on hundreds of examples (Soghomonyan et al., 2021) is the abrupt termination of TGEs by lightning flashes. Simultaneous detection of the time series of particle count rates, energy spectra, and near-surface electric fields, and estimation of distances to lightning flash allows concluding that the area bombarded by RREAs on the earth's surface can reach tens of $km^2$, including millions of gamma rays and rarely electrons and neutrons. These opportunities became feasible, because Aragats facilities, including numerous particle detectors, field meters, and weather stations has been operated without interruptions for many years. Experimental data is available in graphical and numerical types from databases of Cosmic ray division of Yerevan physics institute (http//crd.yearphi.am/ADEI) and from Mendeley data sets (Soghomonyan et al., 2021a, b and Chilingarian and Hovsepyan, 2021).

1. Comparative analysis of 23 - 25 May 2021 thunderstorms

According to the adopted approach of the multivariate correlation analysis, we use as many as possible measurements for characterizing high-energy processes in the atmosphere. The count rate of electron and gamma ray fluxes, as well, as the energy release histograms, are registered by the large spectrometer allowing to recover of differential energy spectra of both charged and neutral fluxes. The basic measurement principle of Aragats solar neutron telescope (ASNT) is described I (Chilingarian et al., 2016 and Chilinarian et al., 2017a). The lightning identification and distance to lightning flash estimation are done by monitoring of disturbances of the near-surface (NS) electric field with the network of EFF-100 electric mills of BOLTEK company and by measurements of the Worldwide lightning location network (WWLLN), one of the nodes of which is located at CRD headquarters premises. Meteorological measurements are made with the DAVIS weather station. Panoramic cameras are used for the monitoring of skies above Aragats. The adopted procedure of the multivariate analysis is implemented for the study of the storm that occurred at the end of May 2021 in the vicinity of the Aragats research station. The main results of this analysis are presented below. On May 23 storm duration was ≈6 hours with more than a hundred nearby lightning flashes. In Fig 1, we present a 16-minute period of the thunderstorm with 2 TGEs terminated by the lightning flashes (distance to lightning flash is denoted by red lines) on the initial stage of development (black, time series of ASNT detector count rates). The duration of each TGE was ≈20 sec, the NS electric field was in the negative domain, the amplitude of the NS field surge caused by the terminating lightning flash was ≈ 50 kV/m (blue time series, also see Table 1). NS electric field shows many episodes of deep negative and deep positive (-20 - +20 kV/m) electric field excursions. The electric field recovery after lightning strikes were very fast (a few seconds). In the first part of the storm numerous attempts to start TGE were registered by the ASNT spectrometer. It is interesting to note that a new TGE started just after lightning terminates the previous one during the electric field recovering stage.  This is evidence of the largely electrified atmosphere when lowering of the potential drop (voltage) by lightning flash did not quench fully electrostatic field and the field very fast returned to the high values exceeding the critical threshold for starting the runaway process. Each new started TGE opens as well the path to the lightning leader as was discussed in (Chilingarian et al., 2017b). Numerous examples of the TGEs preceding lightning flashes are shown in the Mendeley dataset (Soghomonyan et al., 2021a). This dataset and other publication unambiguously show that MeV energy particles are not produced by the lightning bolt, but are multiplied and accelerated in the strong electric fields by the RREA process. For many years we perform monitoring of lightning flashes and particle detector signals with nanosecond accuracy. During these years we did not register any coincidence of thousands of nearby lightning flashes with particle bursts in scintillators, in NaI crystals, and in proportional chambers of neutron monitor (Chilingarian et al., 2019).
There is some kind of interplay between lightning activity and TGE development. When an electric field is very large above the station, multiple RREAs started, however, the combination of the very strong electric field and ionization made by the RREA electrons lead to an early stop of RREA by the lightning flash. Thus, the distant maximum of the lightning activity, as we will see in the next day storm, is preferable for the registration of the long-lasting TGE.

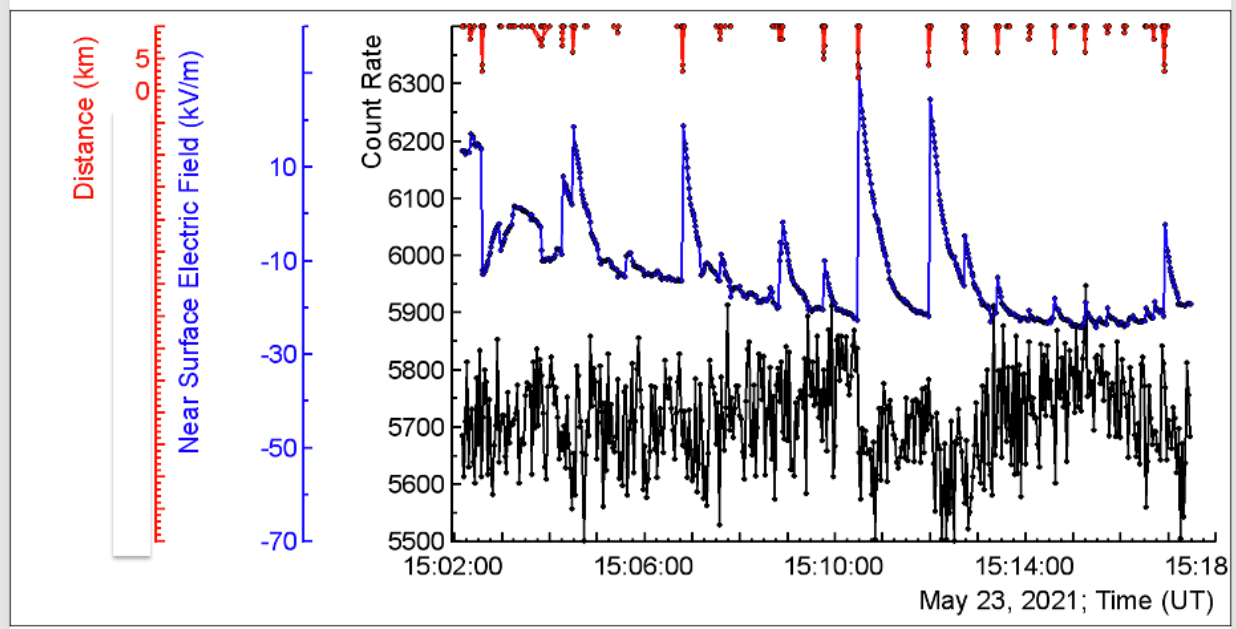

Figure 1. A pattern of the progress of the storm on Aragats on 23 May. Disturbances of the NS electric field (blue), count rate of ASNT spectrometer (black), and distance to lightning (red).

During the second storm that occurred the next day on May 24 and lasted for 3 hours, the electrification of the atmosphere above the station was much smaller. The nearby lightning flashes occurred only during the first hour of the storm terminating a small TGE at the end of its development. During the TGE which occurred 6 minutes later, the nearest lightning flash was at a distance of more than 15 km, and consequently, it did not disturb the smooth evolution of long duration (≈12 minutes) intense TGE (see Table 1 for more details). From the development pattern of this storm (confirmed by the storm that occurred on the next day, May 25) we can conclude, that the RREAs are developing in a huge area in the thundercloud. Near the active lightning region, there are multiple RREAs (and consequently their surface offspring's – TGEs) frequently terminated by lightning flashes and restarted several times at recovering of the strong electric field. If lightning active zone is far (>10 km) from particle detectors we observe long-lasting TGE (from the long-lasting parent RREAs) without termination by lightning flashes. In this case, the scenario with multiple TGE starts and terminations is realized. Thus, the typical TGE on Aragats covered an area where multiple TGEs occurred all terminated by lightning flashes, and a peripheral zone, where lightning activity is reduced allowing long-lasting TGEs to smooth finish.

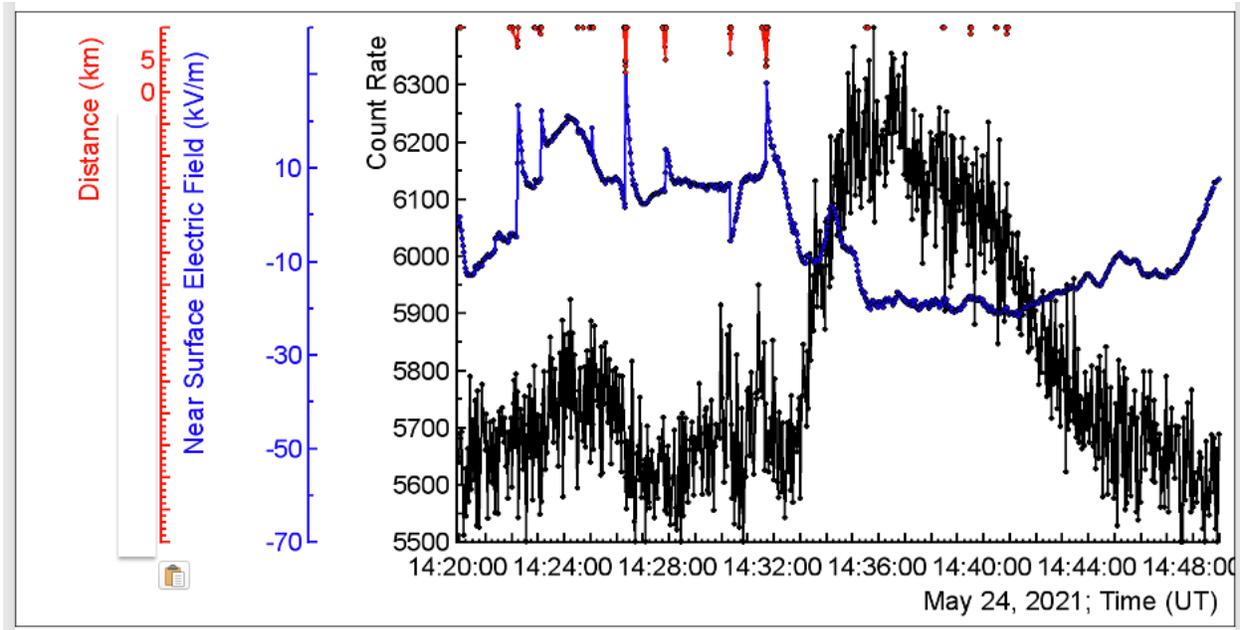

**Figure 2. A pattern of the progress of the storm on Aragats on 24 May. Disturbances of the NS electric field (blue), count rate of ASNT spectrometer (black), and distance to lightning (red).**

On 25 May during a 1-hour long storm, no nearby lightning flashes were registered at all. TGE was lengthy, its duration was ≈18 minutes and particle flux continued both during positive and negative NS electric field, demonstrating that both main scenarios of TGE initiation (with and without emerging LPCR) can be rather smoothly continued. The NS electric field reversal in the lower part of the cloud is an illustration of a decay of a mature LPCR and turning from the second scenario of the RREA initiation (2 dipoles MN-MIRR and MN-LPCR are accelerated electrons) to the first one (only dipole MN-LPCR accelerates electrons, see Fig. 1 in Chilingarian et al., 2021a). Also, it is worth to notice an intense graupel fall during positive NS field, demonstrated contracting of LPCR that is "sitting" on graupels, see Fig.4.

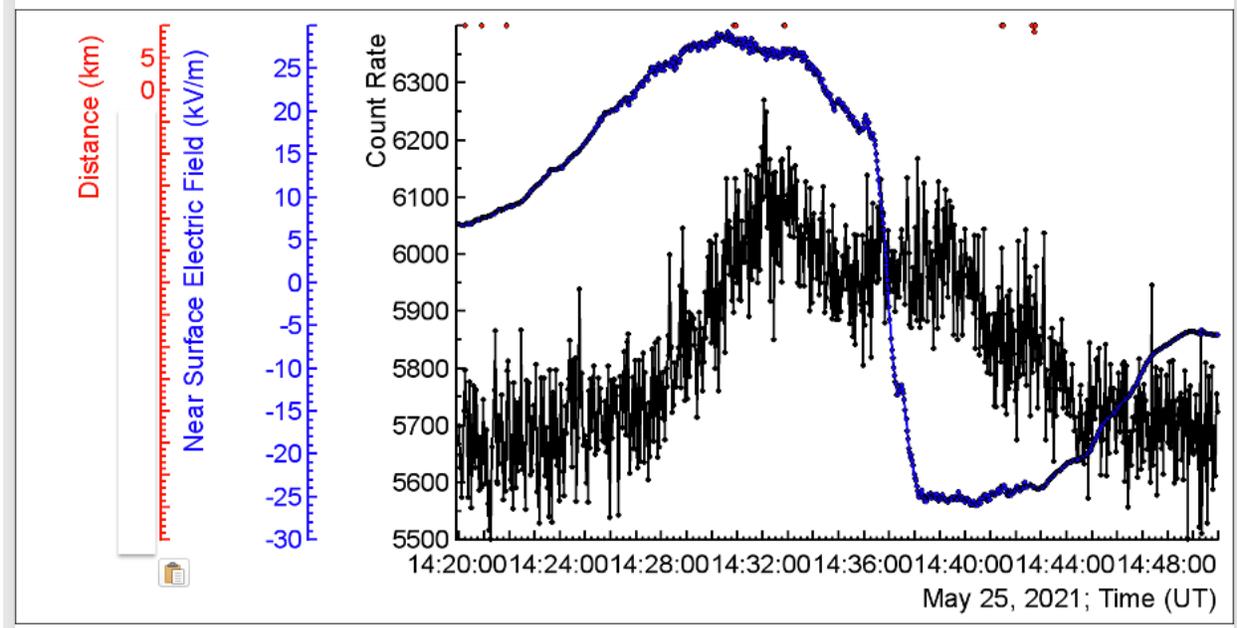

**Figure 3. A pattern of the progress of the storm on Aragats on 25 May. a) The distance to the cloud base. b) Images of the graupel fall. c) Disturbances of the near NS electric field (blue), count rate of ASNT spectrometer (black), and distance to lightning (red).**

In Fig. 4 we show the traces (specks) on the panoramic camera, that monitored skies above Aragats at one frame per minute (shooting frequency enlarged to one frame per second at intense changes of the field of view). The correlation between traces in the camera shots and conical graupel was established by comparing photos of fallen graupel and panoramic camera shots (see Figs 11 and 12 in Chilingarian et al., 2021c). The start and end of the shots with specks coincide with NS electric traversal, i.e., with LPCR contraction.

Sure, LPCR does not screen the MN all the time to induce a positive NS field on the ground as on May 25. There can be several transient cases as demonstrated by Tran and Rakov in their famous paper (Nag & Rakov, 2009) and as we can see in Figs. 1-3. In spite of the graupel being observed also on panoramic shots on 23 and 24 May, the NS electric field was mostly in the negative domain.

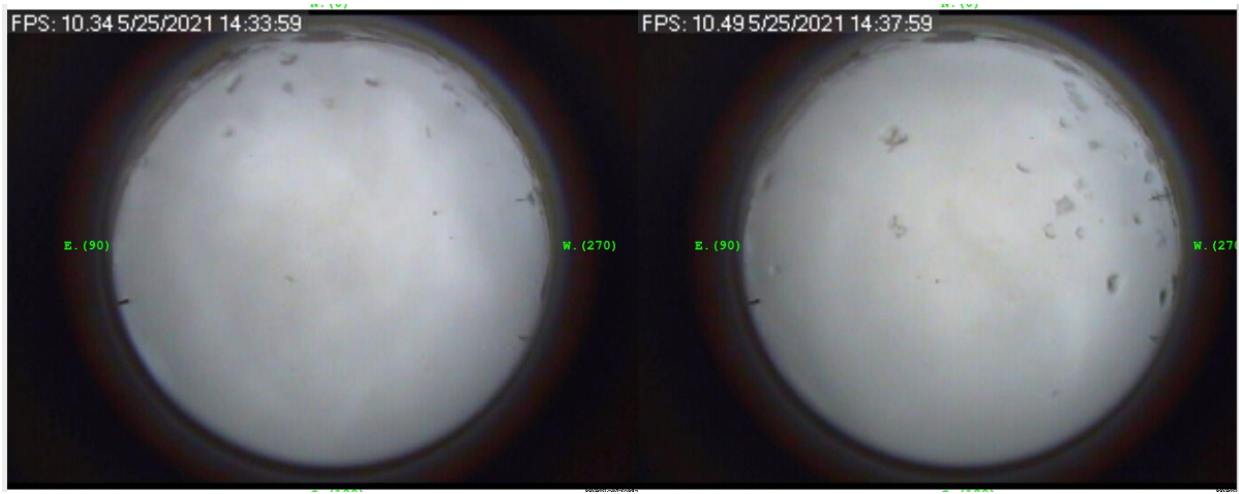

**Figure 4.** Shots of the panoramic camera with characteristic specks on the camera glass, identified as conical graupel fall, the time of start and finish of graupel fall coincides with a mature LPCR contraction.

In Table 1 we summarize characteristics of TGEs registered on Aragats during storms in the end of May 2021. In the first 2 columns we post the TGE start and finish times; in the third the percent of flux enhancement relative to pre-TGE value; in the fourth – mean NS electric field during TGE (on May 25 there were 2 distinct values of NS field – both are posted, separated by a slash); in the fifth column – the mean temperature and distance to cloud base separated by a slash; it the sixth – the time of a terminating of the lightning flash (if any); in the seventh – distance to lightning flash; in the last column - NS electric fields before and after the lightning flash, again separated by a slash.

**Table 1. Parameters of the TGE events and terminating lightning flashes during May storms on Aragats**

| TGE start | TGE Finish | TGE% | Mean Field kV/m | T ©, Dist. Cloud(m) | Time of Flash | Dist. to flash (km) | El. Field change kV/m |
|---|---|---|---|---|---|---|---|
| colspan="8" | TGEs on 23 May,15:00-15:20, totally ≈30 lightning flashes on < 10 km distance |
| 15:10 | 15:19:30 | 2. | -20 | 4.2/300 | 15:19:30 | 1.6 | -21/+31 |
| 15:10:50 | 15:12 | 1.1 | -3 | 4/300 | 15:12 | 5.4 | -21/24 |
| colspan="8" | TGE on 24 May,14:20-14:48, totally ≈15 lightning flashes on < 10 km distance |
| 14:23:01 | 14:26:20 | 4.5 | 17 | 4.5/470 | 14:26:20 | 4.4 | 2/32 |
| 14:32:50 | 14:45:10 | 12.7 | -19 | 1.6/200 | | | |
| colspan="8" | TGE on 25 May,14:20-14:48, totally no lightning flashes on < 10 km distance |
| 14:26:45 | 14:44:45 | 10/8 | +28/-25 | 3.0/240 | | | |

In Fig. 4 we present differential energy spectra measured by a 4 m² area and 60 cm thick scintillation spectrometer ASNT (details on spectrometer operation can be found in Chilingarian et al., 2016, 2017a). Spectra were approximated by the power-law dependence; the power low indices were roughly constant during 4 minutes and equal to ≈2.0 for electron spectra and ≈2.4 – for gamma ray spectra. As it is expected due to large ionization losses of electrons the intensities and maximal energies of the electron flux are smaller than the same parameters of the gamma ray flux. However, the proximity of maximal energies of both TGE species demonstrates that the strong accelerating electric field in the atmosphere was rather low above the earth's surface, lower than the estimate of the cloud base height.

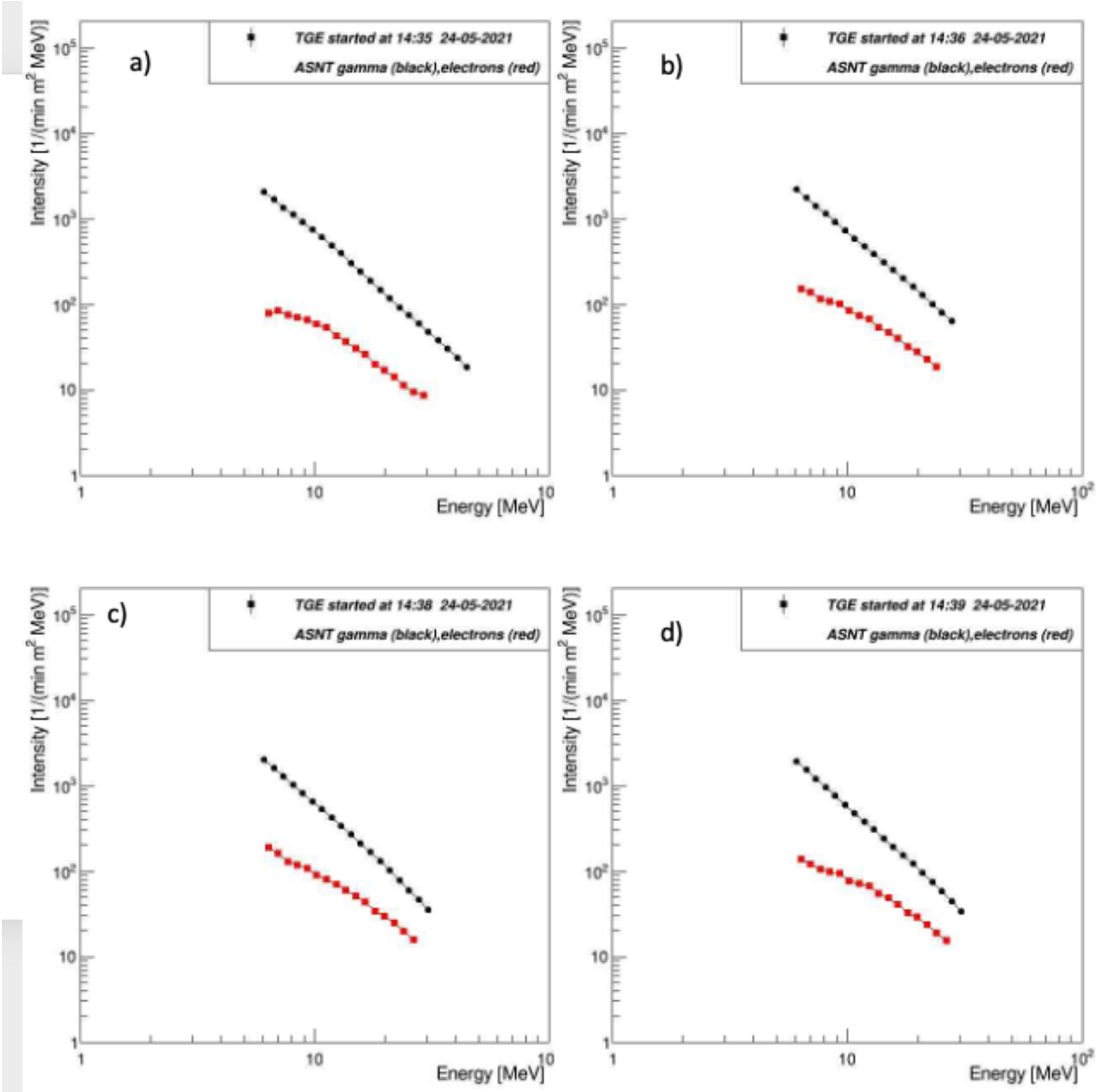

**Figure 5. Differential energy spectra of electrons (red) and gamma rays (black) were registered on May 24, 2021**

In Table 2 we post the parameters of the energy spectra of TGE electrons and gamma rays species for 4 minutes of the maximum TGE flux. In second and third columns – the integral energy spectrum of both fluxes (started from 7 MeV); in fourth – the electron-to-gamma ray ratio; in the fifth and sixth – maximal energies of differential energy spectra for both species; and in the last – an approximate estimate of termination height of the strong accelerating electric field (according to the algorithm from Chilingarian et al., 2021d).

**Table 2. Parameters of gamma ray and electron energy spectra measured on 24 May 2021**

| Date | Integral Spectrum E>7MeV | | | | | |
|---|---|---|---|---|---|---|
| 24 May 2021 | Electron | Gamma Ray | $N_e/N_\gamma$ | $E^e_{max}$ (MeV) | $E^\gamma_{max}$ (MeV) | Est. E.Field height (m) |
| 14:35:00 | 974 | 8090 | 0.12 | 29 | 45 | 125 |
| 14:36:00 | 1530 | 8580 | 0.18 | 24 | 28 | 50 |
| 14:38:00 | 1640 | 7240 | 0.23 | 26 | 31 | 60 |
| 14:39:00 | 1550 | 6660 | 0.23 | 26 | 31 | 60 |

## 2. Justification of the model assumptions

Unfortunately, we are not in a situation to switch on the electron beam with known energy and measured transverse emittance as for target experiments with man-made accelerators. "Thundercloud accelerator" operates in atmospheric electrostatic fields of unknown strength and vertical profile, and avalanche particles reach unknown maximum energies at unknown heights above particle detectors.

We made simulations with models assuming rather simple electric fields profile below 5000-6000m (assumed height of the main negatively charged layer), we test different strengths of the electric field with the spatial extent of 1000 – 2000 m.

Any model is reductant relative to nature, but modelling is the only way to get insight into very complicated physical processes like cloud electrification and RREA. Our goal was to outline the main characteristics of the RREA by exploring possible configurations of the electric field compatible with those sustaining the measured enhancements of particle fluxes. Sure, the RREA simulations with CORSIKA and GEANT4 packages (see simulation details in Chilingarian et al., 2018, and Chilingarian et al., 2021a) were paired with detectors response estimation, which gives the only possibility to compare experimental and simulated data.

By simulations and calculations of the electric field at ground due to the vertical tripole, we conclude that for the TGE physics it is quite enough to consider contributions of main negative and LPCR layers only. The main positively charged layer in the upper part of the cloud does not significantly change the superposition of all induced fields, see Fig.6.

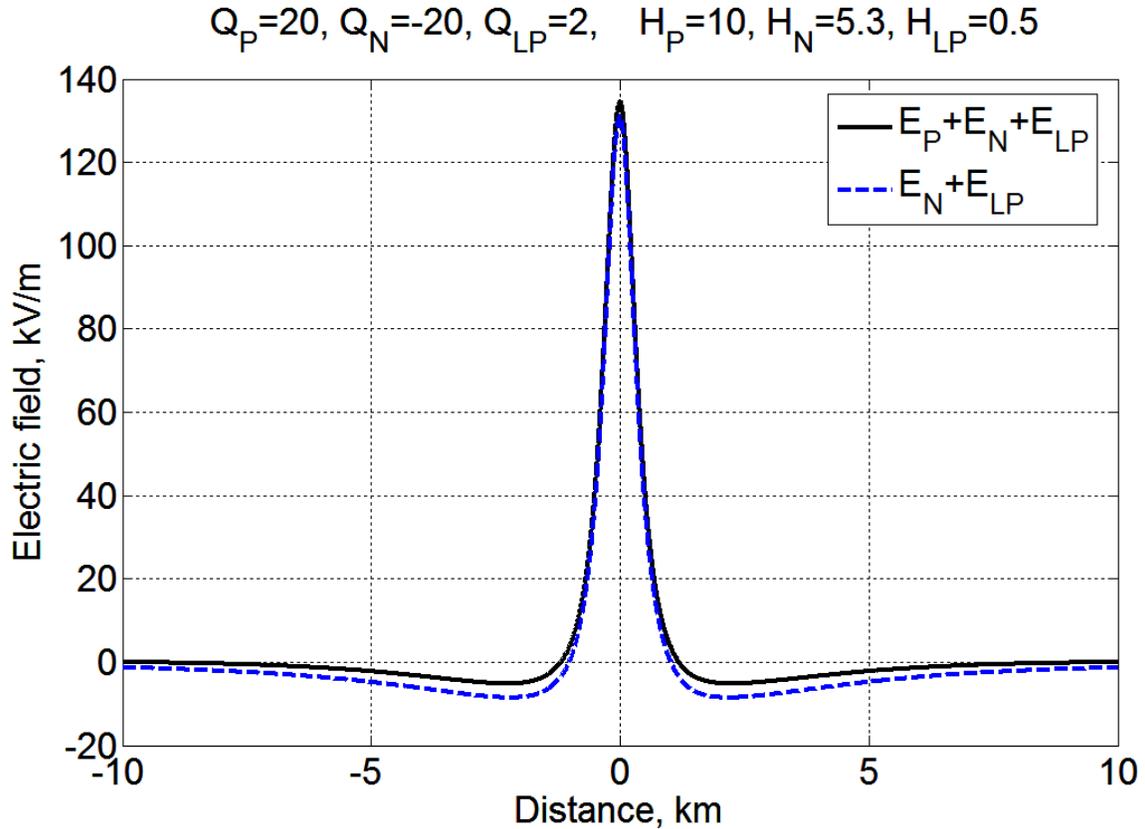

**Figure 6.**

**The electric field at the ground produced by vertical tripole, $Q_P$, $Q_N$, $Q_{LPCR}$, $H_P$, $H_N$, and $H_{LPCR}$ are charges (in Coulombs) and heights (in km) of upper positive, middle negative, and LPCR, respectively (solid black) and by lower dipole only (dashed blue). The abscissa shows the horizontal distance from the tripole axis.**

The electron acceleration in the direction of the earth's surface is due to the electric field between MN and its mirror image induced in the ground. Inside the thundercloud, this field can be significantly increased by an emerging LPCR located below the main negative charge region. The electric field in the gap between the MN layer and LPCR is enhanced by the superposition of $E_{LPCR}$ and $E_{MN}$, while the field below the LPCR will be reduced due to the opposite directions of $E_{LPCR}$ and $E_{MN}$. The maximal intensity (and maximal energy of TGE particles) is observed when the strength of the local electric field in the cloud exceeds the critical threshold and RRE avalanches developed downward. For the operation of the electron accelerator and detection of TGE, we need (1) an electric field above the critical value ($E_{LPCR} + E_{MN}$)> 1.7 kV/m, in the atmosphere at an altitude of 4000–5000 m above the earth's surface, and (2) a large spatial extent of the field (0.5-2 km). Thus, if the MN charge is large enough to induce a strong electric field exceeding the critical value, the RREA can be unleashed and TGE will be large, and gamma ray energies up to 50 MeV will be observed. The near-surface electric field will be in the deep negative domain reaching -20 ÷ 30 kV/m for the largest TGEs. Thus, regardless of the cloud base location, the electric field extends almost down to the earth's surface, and both electrons/positrons and gamma rays can be registered by particle detectors and spectrometers.

When LPCR is mature (the near-surface field is in the positive domain), the fields produced by the MN-mirror and MN-LPCR dipoles are identically directed and their sum can reach very large values inside the cloud.

Sure, TGE can start when LPCR is mature, but after its contraction, only MN sustains a strong electric field (May 25 TGE). Alternatively, in the middle stage of the first scenario, the LPCR is formed and for a few minutes, the near-surface electric field rises and reaches positive values, and then returns again to deep negative values when LPCR is depleted. For both scenarios, the influence of the upper positively charged region is negligible. The electric field between LPCR and its mirror in the earth's surface also does not influence TGE intensity and maximum energy of electrons and gamma rays. After exiting from the strong electric field gamma rays continue their path to the ground, the intensity of gamma ray beam is diminished, but the maximum energy remains rather stable. Electrons are decelerated by the electric field, however, a much more important process frequently stopping the electron flux at all is ionization losses. We estimate the maximum energy of electrons detected by the ASNT to be ≈50 MeV, and the ionization losses of electrons crossing 200 m (from 3400 to 3200), are ≈30 MeV. Thus, TGE electrons with energies above 30 MeV only can reach the ground.

Consequently, the ratio of electrons to gamma rays will be very small, and recovering the electron spectra will be impossible. Sure, gamma rays in the atmosphere will born electrons by the Compton scattering and other processes (with energies much lower than parent gamma rays' energies). However, the percent of these electrons to the parent gamma ray beam in the energy range assessable to our spectrometer (>4 MeV) never exceeds 2.5% for all tested configurations of electric field strengths and elongations, and these electrons will dissipate in the background, see Fig. 7. It is why for all recovered electron energy spectra the $N_e/N_\gamma$ ratio is well above 6% (see Mendeley data set, Soghomonyan et al, 2021b) confirming that it is TGE electrons, and not Compton scattered electrons only.

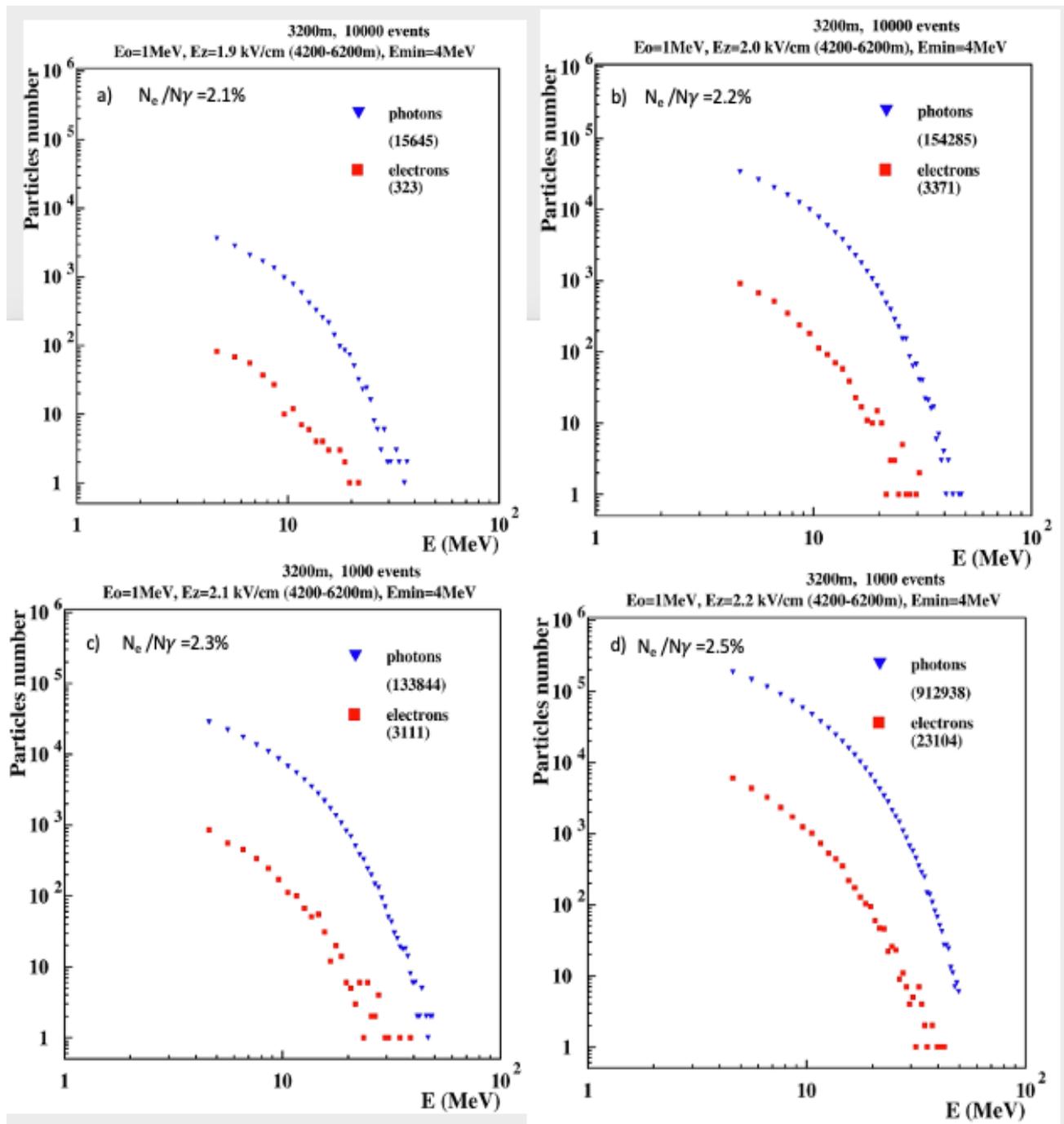

**Figure 7.** Simulated differential energy spectra of electrons and gamma rays reaching the earth's surface from 1000 m (only Compton scattered electrons, no TGE electron can reach the ground from such heights)

**Discussion and conclusions**

Strong electric fields in thunderclouds give rise to RREAs, which end up as TGEs registered by surface particle detectors. Observation of three succeeded each-other TGEs in the end of May 2021 allows to understand interrelation of particle fluxes and atmospheric discharges, energy spectra of electrons and gamma rays discloses the structure and strength of atmospheric electric fields. Three episodes of electron acceleration observed at the end of May 2021 on Aragats demonstrate a rich variability of the electron accelerator operation modes depending on the proximity of the particle detector site to the active storm zone. On 23 May when the storm was just above particle detectors, nearby (at distances 1.6 – 5.4 km) lightning flashes terminate RREA after a few tens of seconds. If the storm active zone is far away from particle detectors (>10 km) the TGE extends 12 and 18 minutes and smoothly terminates when conditions of the atmospheric electric field fail to support RREA. Thus, the RREA can be unleashed in a very large spatial domain around the storm, reaching 10 km and more in radii. Adopting the energy spectra of a TGE registered on May 30, 2018 (Chilingarian et al., 2018) we estimate the total number of gamma rays (with energies above 300 keV) hitting the earth's surface to be $1.3*10^6/m^2min$. Assuming that $\approx 2000$ thunderstorms are active on the globe and that the overall surface of the thunderous atmosphere each moment can be estimated as $2.000 * 100 \text{ km}^2 = 200,000 \text{ km}^2$ (0.04% of the globe surface), we come to an estimate of $1.3*10^{16}$ gamma rays are hitting the earth's surface each second! Proceeding from discussed TGE events and from the vast majority of TGEs registered on Aragats during the last 12 years (more than half-thousand) we confirm statements formulated in (Chilingarian et al., 2017b, see Fig. 1 and Fig. 7) that TGEs are precursors of lightning flashes.

In previous papers we consider two scenarios of TGE initiation:

a) The electric field originated between the main negative and its mirror in the ground (MN-MIRR) forms a lower dipole that accelerates electrons downward;

b) Additional second dipole (MN-LPCR) emerges, which is parallel to the first at least in the most part of the cloud.

LPCR is sitting mainly on precipitation (graupel) that becomes positively charged above $\approx -10C°$. Thus, the falling LPCR elongates the accelerating field until very low heights above the ground and the electric field of the two joint lower dipoles accelerates electrons to high energies due to large potential differences. In the Spring season, almost every thunderstorm and TGE on Aragats was accompanied by a graupel fall which was monitored on a minute time scale. The TGEs occurred on 24-25 May are a good verification of the described above model. We confirm the emergence of the graupel dipole, apparently seen on May 25 (first part of TGE, Fig 3 and 4). It is interesting to notice a minute-long polarity reversal of the NS field (Fig 3), that didn't stop TGE. The NS field fall in the deep negative domain after being in the positive domain for 12 minutes during the first phase of TGE, and afterwards the TGE smoothly continued.

The measured energy spectra and estimates of strong electric field termination heights (Fig.5 and Table 2) also confirmed the low location of the strong electric field in good agreement with the TGE initiation model, and - with simulation results (Chilingarian et.al., 2020 and 2021a).

**Data Availability Statement**

The data for this study are available on the WEB page of the Cosmic Ray Division (CRD) of the Yerevan Physics Institute, http://adei.crd.yerphi.am/adei and in the Mendeley datasets.

**Acknowledgment**

*We thank the staff of the Aragats Space Environmental Center for the operation of the uninterruptable operation of all particle detectors and field meters. Special thanks to T.Karapetyan and B.Sargsyan for tuning and maintaining the ASNT spectrometer. The authors thank S. Soghomonyan for useful discussion, making calculations of the NS electric field (Fig. 6), and helping in preparing the manuscript. The authors acknowledge the support of the Science Committee of the Republic of Armenia (research project № 21AG-1C012), in commissioning particle detectors and in the modernization of technical infrastructure of high-altitude stations. We also acknowledge the support of the Basic Research Program at HSE University, RF for the assistance in running facilities registering atmospheric discharges.*